# Towards Dependable Deep Convolutional Neural Networks (CNNs) with Out-distribution Learning

Mahdieh Abbasi*
Université Laval
mahdieh.abbasi.1@ulaval.ca

Arezoo Rajabi*
Oregon State University
rajabia@oregonstate.edu

Christian Gagné
Université Laval
christian.gagne@gel.ulaval.ca

Rakesh B. Bobba
Oregon State University
rasesh.bobba@oregonstate.edu

*Abstract*—Detection and rejection of adversarial examples in security sensitive and safety-critical systems using deep CNNs is essential. In this paper, we propose an approach to augment CNNs with out-distribution learning in order to reduce misclassification rate by rejecting adversarial examples. We empirically show that our augmented CNNs can either reject or classify correctly most adversarial examples generated using well-known methods ($> 95\%$ for MNIST and $> 75\%$ for CIFAR-10 on average). Furthermore, we achieve this without requiring to train using any specific type of adversarial examples and without sacrificing the accuracy of models on clean samples significantly ($< 4\%$).

## I. INTRODUCTION

Convolutional Neural Networks (CNNs) have become popular due to their high accuracy for image and video analysis. Despite their strong performance, it has been demonstrated that they are highly susceptible to adversarial examples, especially when dealing with high dimensional inputs such as images [30; 29]. An adversarial image is one that has been perturbed with a noise signal designed to fool the CNN. While CNNs confidently misclassify such adversarial examples[1], they are perceptually similar to the original image and easily recognizable by humans (See Figure 1).

Many algorithms for generating adversarial examples have been proposed (*e.g.,* [11; 18; 7; 24]). Those algorithms can broadly be classified as either white-box or black-box attack algorithms. In white-box attacks, an adversary knows the target classifier's exact model parameters and learns adversarial examples over it. In contrast, in black-box attacks, an adversary does not have any knowledge about the target classifier. Therefore, in a black-box attack setting, an adversary learns adversarial examples over a local CNN without access to the target classifier. However, due to the transferability [30] of black-box adversarial images, target CNNs can still be highly vulnerable to them [30; 13].

Furthermore, another serious challenge for CNN-based systems is that when a test sample comes from a different concept or class that is not part of the training set (*i.e.,* in-distribution samples), then CNNs assign them to one of the predefined classes they are trained on, possibly with high confidence. We call such samples as natural (*i.e.,* not synthetically generated) out-distribution samples as they are semantically and statistically different from the in-distribution samples. From a decision-making perspective, this is of great concern as CNNs show confidence that is clearly inappropriate, particularly for security sensitive tasks. This behavior demonstrates that neural networks do over-generalization in some regions that are empty of in-distribution samples (*i.e.,* belonging to out-distribution samples).

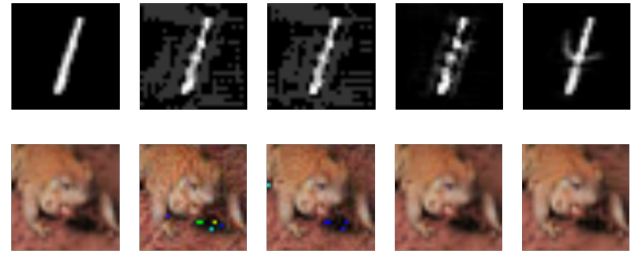

Fig. 1: Adversarial examples for MNIST (first row) and CIFAR-10 (second row). From left to right: original images, T-FGS, FGS, DeepFool, and C&W ($L_2$) adversarial examples

With the aim of addressing these challenges, in this paper we propose to add an additional *dustbin* class containing *(i) natural out-distribution samples* (*i.e.,* natural samples that are statistically and semantically different from in-distribution samples), and *(ii) interpolated in-distribution data* (created by interpolating selected pairs of in-distribution samples from two different classes) to train an augmented CNN. We show that such an augmented CNN has a lower error rate (*i.e.,* misclassification rate) in the presence of adversarial examples because it either correctly classifies adversarial samples or rejects them to the *dustbin* class. Reducing misclassification rate of CNNs is critical for developing secure and dependable CNN-based intelligent systems, particularly in hostile environments. We show that an augmented CNN that is only trained on natural out-distribution samples has the capability of learning more flexible feature space where most of the adversarial examples can be safely separated from in-distribution samples.

In brief, our contributions are as follow:

- We propose a simple yet effective approach to reduce misclassification rate of CNNs for black-box adversarial examples by adding a *dustbin* class to learn a better feature space.
- Unlike previous approaches our approach i) does not need access to adversarial examples for training, ii) does

---

*The first two authors contributed equally

[1]We use the terms adversarial image and adversarial example interchangeably throughout this paper.

not require additional networks (*e.g.,*CNNs, autoencoders *etc.*) to detect adversarial examples, and iii) does not noticeably sacrifice accuracy on clean samples.
- We demonstrate the robustness of our augmented CNNs against well-known attacks (see Sections III and VI). Our augmented CNNs can either reject or classify adversarial examples correctly $> 95\%$ of the time for MNIST and $> 75\%$ of the time for CIFAR-10 on average.

The rest of this paper is organized as follows. In section II we review relevant work. We give a brief introduction to well-known attack methods that we used for evaluating our approach in section III. In section IV we clearly explain our assumptions about the adversary's capabilities and knowledge about the target system. We introduce our method in section V, and evaluate it in section VI. We conclude in Section VII.

## II. RELATED WORK

It has been shown that adversarial examples are easy to generate and they are very successful in fooling even CNNs whose parameters are not known to the adversary (*i.e.,* black-box attacks) [30; 17; 11]. Here, we introduce works that address adversarial examples in image classification for both white-box and black-box attacks.

**Detection and Rejection:** To mitigate the risk of adversarial examples, detection procedures have been proposed for identifying and rejecting adversarial examples [20; 9; 23; 12; 1]. For instance, [20; 12] stated that adversarial samples are actually statistically different from clean samples. So [12] augments the output of a CNN with an extra dustbin label (a.k.a. reject option), then train it on clean training samples and their corresponding adversarial samples (assigned to dustbin) in order to enable CNNs to detect and reject such adversarial samples. However, this assumes that all adversarial sample generation approaches are known and that a diverse set of adversarial samples are accessible for training. Li [20] utilized a cascade classifier to detect adversarial examples, however [5] demonstrated that this method is not always effective. In [9], authors used kernel density estimation in the feature space to identify adversarial samples, with mixed results given that some of the adversarial samples still become entangled with clean samples, which appears difficult to handle by kernel methods. In [22] one or more autoencoder networks are used for detecting adversarial examples by using the reconstruction error for adversarial examples. In [32], authors proposed detecting adversarial examples by looking at the difference between the prediction of three CNNs trained on the original images, modified images with reduced the color bit depth of pixels, and modified images that are spatially. Further, in addition to learning two or more additional CNNs, they need to learn a threshold for detection over adversarial examples which may change from one dataset to another.

**Robust CNNs:** In [27], authors reduce the effectiveness of adversarial samples by using distillation networks. Also, in [26], authors reduce the sensitivity of the CNNs to noise by producing very small gradients at the cost of sacrificing accuracy. However, both those methods are not effective for transferable perturbations and further it has been shown that they can be broken easily [6]. There are couple of works which employed a pre-processing step to denoise inputs in order to classify adversarial examples correctly. For example, in [4], authors used PCA to reduce feature space and remove unnecessary information including noise, and in [8] authors used JPEG compression to remove noise from images and then classify the compressed images. However, both reduce the the accuracy of the CNN and are therefore not practical on large datasets.

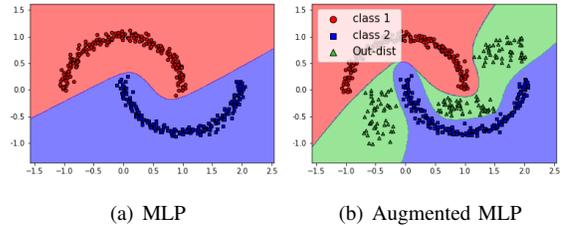

(a) MLP    (b) Augmented MLP

Fig. 2: Two-moons synthetic dataset.(a) MLP trained on only in-distribution samples (b) the augmented MLP trained on both in-distribution and some out-distribution samples

In all previously discussed defense methods, either access to a variety of adversarial examples or learning additional networks is needed. The methods that do not have such overheads sacrifice the accuracy of the model for clean samples by reducing image quality. In contrast, our augmented CNNs are more robust to adversarial examples by rejecting them without needing adversarial samples for training, or using additional networks, or sacrificing accuracy.

## III. PRELIMINARIES

A Convolutional Neural Network (CNN) is constructed by a sequence of blocks, which each consists of convolution, ReLU and pooling layers, followed by a couple of fully connected layers. As the last layer, there is a softmax layer that translates the final fully connected layer outputs to class prediction. In other words, the output of a CNN is treated as a probability vector, where each element denotes conditional probability of each class for a given input $x$. A CNN can be represented by function $Y = F(x,\theta)$ where $x \in [0,1]^d$ is an input image and $\theta$ represents its parameters (*i.e.,* weights and biases).

Generally, an adversarial generation method can either be targeted or untargeted. In targeted attacks, an adversary aims to generate an adversarial sample that makes a victim CNN misclassify it to the victim selected target class (*i.e.,* $\text{argmax}\, F(x+\delta) = y'$, where $y'$ is the targeted class and $\neq y^*$ the actual class). In an untargeted attack, an adversary aims to make the victim CNN to simply misclassify an image to a class other than the true label (*i.e.,* $\text{argmax}\, F(x+\delta) \neq y^*$, where $y^*$ is the true class). Here, we briefly explain some well-known targeted and untargeted attack algorithms.

**Targeted Fast Gradient Sign (T-FGS) [16]:** This targeted attack method tends to modify a clean image $x$ so that the loss function is minimized for a given pair of $(x, y')$, where target class $y'$ is different from the input's true label ($y' \neq y^*$).

To this end, it uses the sign of gradient of loss function as follows:

$$x_{adv} = x - \epsilon.sign(\nabla J(F(x,\theta), y')), \quad (1)$$

where $J(F(x,\theta), y')$ is the loss function and $\epsilon$ as the hyper-parameter controls the amount of distortion. The transferability of T-FGS samples increases by utilizing larger $\epsilon$ at the cost of adding more distortions to the image. Moreover, the untargeted variant of this method called FGS[10] is as follows:

$$x_{adv} = x + \epsilon.sign(\nabla J(F(x,\theta), y^*)). \quad (2)$$

**Iterative Fast Gradient Sign (I-FGS) [18]:** This method actually is an iterative variant of Fast Gradient Sign (also called Projected Gradient Descent (PGD) [21]), where iteratively small amount of FGS's perturbation is added by using a small value for $\epsilon$. To keep the perturbed sample in $\alpha$-neighborhood of $x$, the achieved adversary sample in each iteration should be clipped.

$$\begin{aligned} x_{adv}^0 &= x \\ x_{adv}^{k+1} &= clip_{x,\alpha}\{x_{adv}^k + \epsilon.sign(\nabla J(F(x_{adv}^k, \theta), y^*))\}, \end{aligned} \quad (3)$$

Compared to FGS, I-FGS generates more optimal distortions.
**DeepFool [24]:** This algorithm is an iterative but fast approach for creation of untargeted attacks with very small amount of perturbations. Indeed, DeepFool generates sub-optimal perturbation for each sample where the perturbation is designed to transfer the clean sample across its nearest decision boundary.
**Carlini Attack (C&W) [7]:** Unlike previous proposed methods which find the adversarial examples over loss functions of CNN, this method defines a different objective function which tends to optimize misclassification as follows:

$$f(x') = \max(\max\{Z(x')_{y'} - Z(x')_{y^*}\}, -\kappa) \quad (4)$$

Here $Z(x)$ is the output of last fully connected (before softmax) layer and $x'$ is perturbed image $x$. Also $\kappa$ denotes confidence parameter. A larger value for $\kappa$ leads the CNN to misclassify the input more confidently, however it also makes finding adversarial examples satisfying the condition (having high misclassification confidence) difficult.

## IV. ADVERSARY MODEL

It has been shown that it is easy to break detection and rejection approaches when an adversary has access to details of the classifiers as well as detectors (*i.e.,* white-box attacks) [14; 7]. However most real-world classifiers (*e.g.,* online classifier services) do not publish the details of their classifiers or their defenses. Nevertheless, it has been demonstrated that adversarial examples are highly transferable to other CNNs even when their used parameters and hyper-parameters are unknown to an adversary (*i.e.,* black-box attacks) since CNNs have been shown to share similar boundaries [31]. Some recent works such as feature-squeezing [32] consider another setting that we refer to as gray-box attacks, where the details of the target CNN (*i.e.,* parameters and hyper-parameters) are known to the adversary but the underlying defenses (*i.e.,* detectors) are undisclosed.

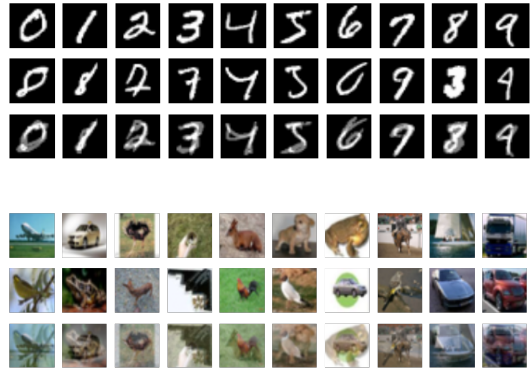

Fig. 3: Interpolated samples for MNIST and CIFAR10. Third row represents the interpolated samples that are composed of images from first (source) and second rows (target).

In this paper, we focus on black-box adversarial attacks as they can readily be utilized to target any unknown CNN. To create adversarial examples that can be transferred to fool target CNNs, an adversary needs to train local classifier(s). Thus, we assume that she has access to some clean samples (excluding dustbin samples) to learn a classifier and she is not aware of the utilized defenses and classifiers' hyper parameters. Moreover, we assume that the adversary has enough resources to generate any type of adversarial attacks.

## V. PROPOSED METHOD

It has been argued that a central element explaining the success of deep neural networks is their capacity to learn distributed representations [2]. Indeed, this allows such models to perform well in the regions that are only sparsely sampled in the training set, specially in a very high dimension space. However, neural networks make totally arbitrary decisions in the regions that are outside of the distribution of the learned concepts, leading to over-generalization. For instance, the classification regions achieved from two MLPs on the well-known two-moons dataset are illustrated in Fig. 2. Adding a *dustbin* class by including some out of distribution samples to the training set leads to more accurate decision regions, and therefore the over-generalization effect in these out-distribution regions can be reduced. In this paper, we leverage this to reduce over-generalization by training an augmented CNN with a *dustbin* class for which available natural out-distribution images are used as training samples. We then investigate the effect of this over-generalization reduction on five known powerful adversarial attacks (discussed in Section III). Indeed, as seen in Figure 4, we find that the over-generalization reduction leads to a more expressive feature space where all natural out-distribution samples along with many black-box adversarial examples are separated from in-distribution samples to be classified as belonging to the *dustbin* class. Further, some adversarial instances are even placed very close to their corresponding true class, leading the augmented CNNs to classify them correctly.
**Interpolated Data:** To increase the rejection power of the augmented CNN on adversarial examples generated using powerful attack algorithms that generate small amount of smart adversarial perturbations (e.g. DeepFool and C&W attacks), we utilize some interpolated images that are generated from in-distribution samples as training sample for *dustbin* class. Our intuition is that an adversarial example simultaneously contains the relevant features from two classes (*i.e.,* the target class and the true class), where the features of the fooling class are hardly visible while the features related to the true class can be recognizable for human observers. Accordingly, we interpolate some images from in-distribution samples and ensure each resulting image has the features from two different classes. To create interpolated data, we simply add two weighted images together as follows:

$$I_{c_i,c_j,\alpha} = \alpha I_{c_i} + (1-\alpha) I_{c_j} \qquad (5)$$

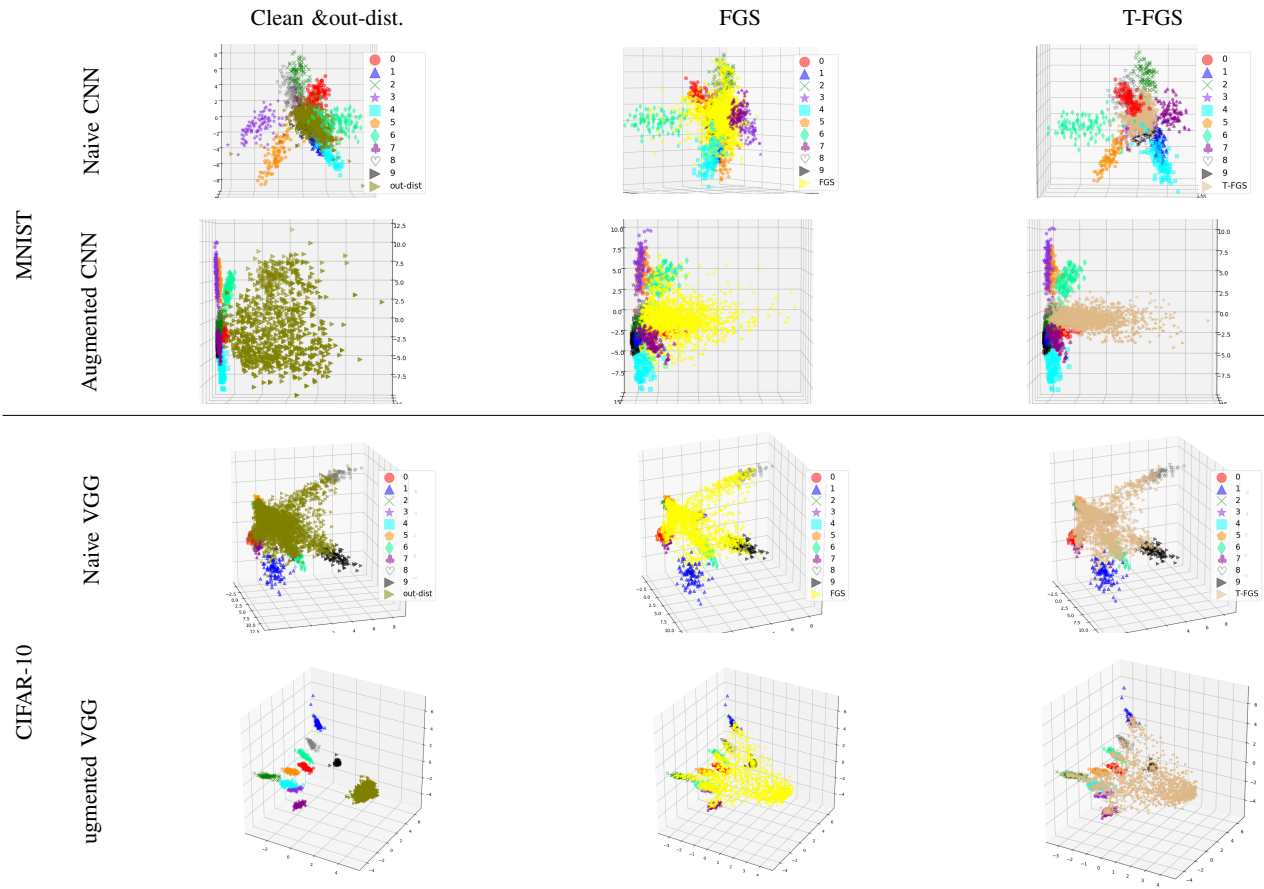

Fig. 4: Visualization of some randomly selected test samples and their corresponding adversaries (FGS and T-FGS) in the feature spaces (the penultimate layer) learned by a naive CNN and an augmented CNN. To produce and manipulate the 3D plots refer to https://github.com/mahdaneh/Out-distribution-learning_FSvisulization

Here, $I_{c_i}$ and $I_{c_i}$ are images from source and target classes. Also, $\alpha$ is a mixing-ratio parameter.

**Pair Selection:** For generating interpolated data, considering all possible combinations of classes for all training samples of a large-scale dataset and using them for training can be computationally expensive. Therefore, we randomly select a subset of samples from each source class (choosing 1500 samples out of 5000 samples in our simulation). Then, for each selected source sample, its nearest neighbor which is located in another class is found. To do so, we exploit the penultimate layer of a naive CNN as a feature extractor [3; 2] so that we can accurately measure the similarity between images in the feature space rather than in the pixel space. Furthermore, finding nearest neighbors in feature space is considerably faster than finding them in pixel space due to noticeably smaller dimensionality of feature space (*e.g.,*512 vs 3072 for CIFAR). Finally, using $\alpha = 0.5$, we generate an interpolated sample for each pair of source image and its target image. , we can say interpolated image samples belong equally to the source and target classes. Larger (or smaller) values for $\alpha$ leads the accuracy of CNNs to drop and false positive rate to increase, since interpolated data would be very similar to clean samples. Figure 3 shows some interpolated samples along with their source and target images.

**Out-Distribution Samples:** To choose out-distribution samples, we considered images with objects that are totally different from objects in training and test set images. The number of out-distribution samples added to dustbin class depends on the number of clean examples and number of classes. A very large dustbin class compared to other classes leads the generalization error to increase and a small ones does not improve the resiliency of our augmented CNN. For our evaluation, we increased the training set by 50%.

## VI. EVALUATION

Using MNIST [19] and CIFAR-10 [15], we empirically evaluated our proposed method against black-box adversarial samples which are generated using five different well-known attack algorithms. Specifically, we compared the error rates of augmented CNNs with that of naive CNNs for adversarial examples. Here the error rate represents the percentage of wrong decisions made by CNNs for adversarial examples. That is, where the adversarial examples are neither rejected nor correctly classified. Therefore, we compute the error rate as $1 - (accurate\ calssification\ rate + rejection\ rate)$. We also compared the accuracy of the naive and augmented CNNs for clean samples.

**MNIST with NotMNIST**: MNIST is a popular machine learning dataset that contains labeled gray scale images, where each image holds a hand-written digit. As out-distribution samples for MNIST, we considered NotMNIST dataset[2] that consists of 18,724 letters A-J printed with different font styles. Images of both MNIST and NotMNIST datasets have the same size images ($28 \times 28$ pixels). We used a CNN named cuda-convnet that has three convolution layers with 32, 32, and 64 filters of $5 \times 5$, respectively, and one Fully Connected (FC) layer with softmax activation function[3]. In addition, dropout with $p = 0.5$ is used at the FC layer for regularization. To train an augmented version of cuda-convnet, we utilized a training set comprising $50K$ MNIST training samples as in-distribution data and $10K$ randomly selected samples from NotMNIST dataset along with $15K$ interpolated samples (see SectionV) generated from MNIST training samples as out-distribution samples. The remaining samples from NotMNIST ($\approx$8K) in conjugation with MNIST test samples are considered for evaluating the augmented CNN.

**CIFAR-10 with CIFAR-100**: Training and test sets of CIFAR-10 contain 50K and 10K RGB images ($32 \times 32$ pixels each). As out-distribution samples for CIFAR-10, we consider CIFAR-100 dataset. To reduce a conceptual overlap between the labels from CIFAR-10 and CIFAR-100, we ignore super-classes of CIFAR-100 that are conceptually similar to CIFAR-10 classes. So, vehicle 1, vehicle 2, medium-sized mammals, small mammals, and large carnivores are excluded from CIFAR-100. Note, all the images are scaled to $[0, 1]$, then normalized by mean subtraction over the CIFAR-10 training set. For CIFAR-10, we chose VGG-16 [28] architecture that has 13 convolution layers with filter size 3x3 and three FC layers. To train an augmented VGG-16, $15K$ randomly selected samples from the non-overlapped version of CIFAR-100 along with $15k$ interpolated samples from CIFAR-10 training set (labeled as dustbin class) are appended to CIFAR-10 training set.

[2]Available at http://yaroslavvb.blogspot.ca/2011/09/notmnist-dataset.html.
[3]To read more about the configuration of this CNN, the readers should refer to https://github.com/dnouri/cuda-convnet/blob/master/example-layers/layers-18pct.cfg

**Evaluation:** As adversarial examples are transferable to other CNNs [30; 25], we learned adversarial examples using T-FGS, FGS and DeepFool approaches over independently trained instances of cuda-convnet CNN for MNIST and VGG16 for CIFAR-10. All correctly classified test samples from each dataset are regarded for adversarial example generation by each of the aforementioned attack algorithms. For FGS and T-FGS attacks, we utilized $\epsilon = 0.2$ for MNIST and $\epsilon = 0.03$ for CIFAR-10. For I-FGS attacks, $\epsilon = 0.003$ along with $\alpha = 0.05$ for CIFAR-10 and $\epsilon = 0.02$ along with $\alpha = 0.2$ for MNIST are used. To generate targeted Carlini attack (called C&W) [7], we used the authors' github code. Due to large time complexity of C&W, we considered 100 randomly selected images for each dataset. For each selected image, as was done in previous work [32], two targeted adversarial samples are generated, where the target classes are the least likely and most likely classes according to the predictions provided by the underlying CNN. Thus, in total 200 C&W adversarial examples are generated per dataset. To increase tranferability of C&W, we utilized $\kappa = 20$ for MNIST and $\kappa = 10$ for CIFAR-10 (see Fig.1 for some adversarial examples).

The results are shown in Table I, where, "Acc." denotes the accuracy of models in classification of adversarial examples, and column "Rej." represents the percentage of adversarial samples rejected by the augmented CNNs (*i.e.,*classified as *dustbin*). Finally, "Err" ($= 1 - ("Acc." + "Rej.")$) shows the misclassification rate (i.e. the percentage of wrong decisions). As seen in Table I, augmented CNN maintains the same accuracy as naive CNN for clean MNIST test samples, while accuracy of augmented VGG on clean CIFAR-10 test samples drops by $\approx 4\%$ points when compared to naive VGG.

Even though our augmented CNNs are not trained on adversarial inputs generated using any specific algorithm, they were able to reject (classify as *dustbin*) most of the adversarial inputs generated using 5 well-known attack algorithms proposed in literature, and even correctly classify a small portion of them. For example, our augmented CNN either rejected or correctly classified almost 96% of the adversarial examples on average, while the Naive CNN misclassified 79% of them. Similarly, for CIFAR-10, the misclassification rate reduces to 24% for augmented VGG when the Naive VGG misclassifies 55% of the adversarial sample on average. Further, as shown in this table, adding interpolated data to dustbin class improved the adversarial sample detection rate over just using natural out-distribution data.

Compared to the most recent defense approach [32], we achieved a lower error rate on average over FGS, I-FGS and DeepFool attacks against CIFAR-10 and MNIST. For FGS attacks against MNIST, our error rate was 1% more than what was reported in [32] (0%). However, unlike the approach in [32] we achieved the low error rate (1%) without having to learn additional CNNs. For C&W attacks though, our method had higher error rate (15.1% for MNIST and 21.5% for CIFAR-10) compared to what was reported in [32] (0%). However, apart from not having to learn additional CNNs, we also used a higher[4] value for $\kappa$ ($\kappa = 20$ for MNIST; $\kappa = 10$ for CIFAR-10) to generate more transferable attacks. For example, for $\kappa = 0$ our naive MNIST CNN itself was able to correctly classify 97% of C&W adversarial samples as compared to only 22.49% of C&W adversarial samples for $\kappa = 20$.

| Training set: <in-dist, out-dist.> | <MNIST, —> | <MNIST, NotMNIST> | <MNIST, NotMNIST+intrpl.> | <CIFAR-10, —> | <CIFAR-10, CIFAR100> | <CIFAR-10, CIFAR100+intrpl.> |
|---|---|---|---|---|---|---|
| Model | Naive CNN | Augmented CNN | Augmented CNN | Naive VGG | Augmented VGG | Augmented VGG |
| In-dist. test Acc. | **99.5** | 99.47 | 99.48 | **90.53** | 88.58 | 86.65 |
| Out-dist. test Rej. | - | 99.96 | **99.98** | - | 95.36 | **96.21** |
| FGS Acc | 35.14 | 19.15 | 0.35 | 36.16 | 27.65 | 23.94 |
| FGS Rej | – | 65.19 | 99.59 | - | 38.94 | 49.23 |
| FGS Err | 64.86 | 15.66 | **0.06** | 63.84 | 33.41 | **26.83** |
| I-FGS Acc | 16.37 | 30.97 | 0.0 | 50.34 | 45.98 | 41.92 |
| I-FGS Rej | - | 27.08 | 100 | - | 18.57 | 25.88 |
| I-FGS Err | 83.63 | 41.95 | **0.0** | 49.66 | 35.45 | **32.2** |
| T-FGS Acc | 19.99 | 1.17 | 0.0 | 36.24 | 27.06 | 24.2 |
| T-FGS Rej | - | 95.92 | 100 | - | 40.54 | 50.77 |
| T-FGS Err | 80.01 | 2.91 | **0.0** | 63.76 | 32.4 | **25.03** |
| DeepFool Acc | 1.89 | 11.45 | 5.37 | 56.82 | 45.63 | 42.31 |
| DeepFool Rej | - | 4.72 | 89.84 | - | 31.0 | 38.86 |
| DeepFool Err | 98.11 | 83.83 | **4.8** | 43.18 | 23.37 | **18.83** |
| C&W ($L_2$) Acc | 22.49 | 27.5 | 7.5 | 42.5 | 46.5 | 39 |
| C&W ($L_2$) Rej | - | 5.99 | 77.49 | - | 18.5 | 39.5 |
| C&W ($L_2$) Err | 77.51 | 66.51 | **15.01** | 57.5 | 35 | **21.5** |
| Average Error rate | 80.82 | 42.17 | **3.97** | 55.59 | 31.92 | **24.88** |

TABLE I: Performance on black-box adversaries attacks. "Acc." corresponds to accuracy (the rate of correctly classified samples), "Rej." is the rejection rate, while "Err." is the misclassification rate All results reported are percentages (%).

Moreover, to show how adding a *dustbin* class can effect the feature space learned by a CNN, we plotted the feature spaces obtained from the last convolutional layers of a naive CNN and its corresponding augmented CNN for clean test samples and their corresponding black-box FGS and T-FGS adversarial examples. In Figure 4, one can see that adversarial examples are located very close to the in-distribution samples in the feature space of a Naive CNN. While our augmented CNN, which is trained on both in-distribution and natural out-distribution samples, is able to disentangle the adversarial samples from in-distribution samples in its feature space even though it was not trained on adversarial examples generated using any specific attack approach.

## VII. CONCLUSION

In this paper we empirically demonstrate how CNNs augmented with out-distribution learning can either reject or correctly classify adversarial examples. Unlike previously proposed approaches our out-distribution learning approach does not rely on having access to adversarial examples which is useful as the approaches for generating such examples rapidly change and evolve. However, our approach is not foolproof and we aim to assess our method with other attack models and on larger datasets. Also, we plan to assess our approach against white-box attacks and explore how it can be combined with other defenses to further reduce the error rate.

## ACKNOWLEDGEMENT

This work was made possible in part through funding from NSERC-Canada, U.S. National Science Foundation (1329681), MITACS, and E Machine Learning Inc. Computational resources were provided by Compute Canada/Calcul Québec and by a GPU grant from NVIDIA. The authors would like to thank Marc-Andrè Gardner, Alhussein Fawzi for their constructive feedback. We also are thankful to the anonymous reviewers for their helpful comments for improving the final version of the paper.

---

[4]This is assuming that the authors of [32] used $\kappa$ value of 0 for C&W attacks as they did not explicitly report the $\kappa$ value other than saying they used the original implementation of C&W which sets $\kappa = 0$ by default. A lower value of $\kappa$ leads to lower transferability.

## REFERENCES

[1] M. Abbasi and C. Gagné. Robustness to adversarial examples through an ensemble of specialists. *arXiv preprint arXiv:1702.06856*, 2017.

[2] Y. Bengio. Learning deep architectures for ai. *Foundations and trends® in Machine Learning*, 2(1):1–127, 2009.

[3] Y. Bengio, G. Mesnil, Y. Dauphin, and S. Rifai. Better mixing via deep representations. In *International Conference on Machine Learning*, pages 552–560, 2013.

[4] A. N. Bhagoji, D. Cullina, C. Sitawarin, and P. Mittal. âenhancing robustness of machine learning systems via data transformations, 2017.

[5] N. Carlini and D. Wagner. Adversarial examples are not easily detected: Bypassing ten detection methods. *Proceedings of the 10th ACM Workshop on Artificial Intelligence and Security*.

[6] N. Carlini and D. Wagner. Defensive distillation is not robust to adversarial examples. *arXiv preprint arXiv:1607.04311*, 2016.

[7] N. Carlini and D. Wagner. Towards evaluating the robustness of neural networks. In *Security and Privacy (SP), 2017 IEEE Symposium on*, pages 39–57. IEEE, 2017.

[8] N. Das, M. Shanbhogue, S.-T. Chen, F. Hohman, L. Chen, M. E. Kounavis, and D. H. Chau. Keeping the bad guys out: Protecting and vaccinating deep learning with jpeg compression. *arXiv preprint arXiv:1705.02900*, 2017.

[9] R. Feinman, R. R. Curtin, S. Shintre, and A. B. Gardner. Detecting adversarial samples from artifacts. *arXiv preprint arXiv:1703.00410*, 2017.

[10] I. J. Goodfellow, J. Shlens, and C. Szegedy. Explaining and harnessing adversarial examples. *arXiv preprint arXiv:1412.6572*, 2014.

[11] I. J. Goodfellow, J. Shlens, and C. Szegedy. Explaining and harnessing adversarial examples. *International Conference on Learning Representations*, 2015.

[12] K. Grosse, P. Manoharan, N. Papernot, M. Backes, and P. McDaniel. On the (statistical) detection of adversarial examples. *arXiv preprint arXiv:1702.06280*, 2017.

[13] K. He, X. Zhang, S. Ren, and J. Sun. Deep residual learning for image recognition. *arXiv preprint arXiv:1512.03385*, 2015.
[14] W. He, J. Wei, X. Chen, N. Carlini, and D. Song. Adversarial example defenses: Ensembles of weak defenses are not strong. *USENIX Workshop on Offensive Technologies*, 2017.
[15] A. Krizhevsky and G. Hinton. Learning multiple layers of features from tiny images. 2009.
[16] A. Kurakin, I. Goodfellow, and S. Bengio. Adversarial examples in the physical world. *arXiv preprint arXiv:1607.02533*, 2016.
[17] A. Kurakin, I. Goodfellow, and S. Bengio. Adversarial machine learning at scale. *arXiv preprint arXiv:1611.01236*, 2016.
[18] A. Kurakin, I. Goodfellow, and S. Bengio. Adversarial examples in the physical world. *International Conference on Learning Representations*, 2017.
[19] Y. LeCun. The mnist database of handwritten digits. *http://yann. lecun. com/exdb/mnist/*, 1998.
[20] X. Li and F. Li. Adversarial examples detection in deep networks with convolutional filter statistics. *CoRR, abs/1612.07767*, 7, 2016.
[21] A. Madry, A. Makelov, L. Schmidt, D. Tsipras, and A. Vladu. Towards deep learning models resistant to adversarial attacks. *arXiv preprint arXiv:1706.06083*, 2017.
[22] D. Meng and H. Chen. Magnet: a two-pronged defense against adversarial examples. In *Proceedings of the 2017 ACM SIGSAC Conference on Computer and Communications Security*, pages 135–147. ACM, 2017.
[23] J. H. Metzen, T. Genewein, V. Fischer, and B. Bischoff. On detecting adversarial perturbations. *5th International Conference on Learning Representations (ICLR)*, 2017.
[24] S. M. Moosavi Dezfooli, A. Fawzi, and P. Frossard. Deepfool: a simple and accurate method to fool deep neural networks. In *Proceedings of 2016 IEEE Conference on Computer Vision and Pattern Recognition (CVPR)*, number EPFL-CONF-218057, 2016.
[25] N. Papernot, P. McDaniel, I. Goodfellow, S. Jha, Z. B. Celik, and A. Swami. Practical black-box attacks against machine learning. In *Proceedings of the 2017 ACM on Asia Conference on Computer and Communications Security*, pages 506–519. ACM, 2017.
[26] N. Papernot, P. McDaniel, A. Sinha, and M. Wellman. Towards the science of security and privacy in machine learning. *IEEE European Symposium on Security and Privacy*, 2018.
[27] N. Papernot, P. McDaniel, X. Wu, S. Jha, and A. Swami. Distillation as a defense to adversarial perturbations against deep neural networks. *Proceedings of the 37th IEEE Symposium on Security and Privacy*, 2016.
[28] K. Simonyan and A. Zisserman. Very deep convolutional networks for large-scale image recognition. *arXiv preprint arXiv:1409.1556*, 2014.
[29] C. Szegedy, W. Liu, Y. Jia, P. Sermanet, S. Reed, D. Anguelov, D. Erhan, V. Vanhoucke, and A. Rabinovich. Going deeper with convolutions. In *Proceedings of the IEEE Conference on Computer Vision and Pattern Recognition*, pages 1–9, 2015.
[30] C. Szegedy, W. Zaremba, I. Sutskever, J. Bruna, D. Erhan, I. Goodfellow, and R. Fergus. Intriguing properties of neural networks. 2014.
[31] F. Tramèr, N. Papernot, I. Goodfellow, D. Boneh, and P. McDaniel. The space of transferable adversarial examples. *arXiv preprint arXiv:1704.03453*, 2017.
[32] W. Xu, D. Evans, and Y. Qi. Feature squeezing: Detecting adversarial examples in deep neural networks. *Network and Distributed System Security Symposium*, 2018.